\documentclass[journal]{IEEEtran}

\usepackage{xcolor,soul,framed} %,caption
\usepackage{hyperref}
\colorlet{shadecolor}{yellow}
\usepackage[pdftex]{graphicx}
\graphicspath{{../pdf/}{../jpeg/}}
\DeclareGraphicsExtensions{.pdf,.jpeg,.png}

\usepackage[cmex10]{amsmath}
\usepackage{array}
\usepackage{mdwmath}
\usepackage{mdwtab}
\usepackage{eqparbox}
\usepackage{url}

\begin{document}
\bstctlcite{IEEEexample:BSTcontrol}
    \title{Security and Privacy Issues of Federated Learning}
  \author{Jahid~Hasan~\IEEEmembership{}
      ~\IEEEmembership{}\\
      % Ignacio~Ramos,~\IEEEmembership{Student Member,~IEEE,}
      % Erez Falkenstein,~\IEEEmembership{Student Member,~IEEE,}
      % and~Zoya~Popovi\'c,~\IEEEmembership{Fellow,~IEEE}% <-this % stops a space

  % \thanks{Manuscript received July 10, 2012. \hl{This paper is an expanded paper from the IEEE MTT-S Int. Microwave Symposium held on June 17-22, 2012 in Montreal, Canada.} This work was funded in part by the Office of Naval Research under the Defense Advanced Research Projects Agency (DARPA) Microscale Power Conversion (MPC) Program under Grant N00014-11-1-0931, and in part by the Advanced Research Projects Agency-Energy (ARPA-E), U.S. Department of Energy, under Award Number DE-AR0000216.}
  % \thanks{M. Roberg is with TriQuint Semiconductor, 500 West Renner Road Richardson, TX 75080 USA (e-mail: michael.roberg@tqs.com).}% <-this % stops a space
  \thanks{Jahid Hasan, Department of Computer Science, Iowa State University, Ames, IA 50011. E-mail: {jhasan}@iastate.edu}
 }

% The paper headers
% \markboth{IEEE TRANSACTIONS ON KNOWLEDGE AND DATA ENGINEERING, (NOVEMBER 2020)
% }{Jahid and Cai: High-Efficiency Diode and Transistor Rectifiers}

% ====================================================================
\maketitle

% === ABSTRACT ====================================================================
% =================================================================================
\begin{abstract}
Federated Learning (FL) has emerged as a promising approach to address data privacy and confidentiality concerns by allowing multiple participants to construct a shared model without centralizing sensitive data. However, this decentralized paradigm introduces new security challenges, necessitating a comprehensive identification and classification of potential risks to ensure FL's security guarantees. This paper presents a comprehensive taxonomy of security and privacy challenges in Federated Learning (FL) across various machine learning models, including large language models. We specifically categorize attacks performed by the aggregator and participants, focusing on poisoning attacks, backdoor attacks, membership inference attacks, generative adversarial network (GAN) based attacks, and differential privacy attacks. Additionally, we propose new directions for future research, seeking innovative solutions to fortify FL systems against emerging security risks and uphold sensitive data confidentiality in distributed learning environments.
\end{abstract}

% === KEYWORDS ====================================================================
% =================================================================================

\IEEEpeerreviewmaketitle

% === I. INTRODUCTION =============================================================
% =================================================================================
\section{Introduction}

\IEEEPARstart{I}{n} recent years, Federated Learning (FL) has emerged as a promising and transformative paradigm for addressing data privacy and confidentiality concerns in the realm of machine learning. Unlike traditional centralized approaches, FL allows multiple participants to collaboratively construct a shared model without the need to centralize sensitive data. By empowering individual devices or entities to train models locally and share only model updates with a central aggregator, FL offers a privacy-preserving alternative for harnessing the collective intelligence of distributed data sources.

The decentralized nature of FL brings forth a new set of security challenges that demand rigorous investigation and mitigation strategies. Federated Learning (FL) has emerged as a revolutionary approach to uphold user privacy by distributing data from the central server to individual devices, empowering various domains with sensitive data and diverse characteristics to benefit from AI advancements. This novel paradigm gained prominence for two primary reasons\cite{mothukuri2021survey}: Firstly, it addresses the challenge of inadequate centralized data access in traditional machine learning. Due to direct access restrictions, certain data cannot be stored on the central server. And secondly, it ensures data privacy protection by utilizing local data from edge devices, such as clients, instead of transmitting sensitive information to the server. This way, the network's asynchronous communication comes into play, preserving the confidentiality of data. By safeguarding data privacy, federated learning allows for the efficient utilization of AI benefits across multiple domains through machine learning models. As participants operate independently, there is an inherent risk of potential security threats that may undermine the integrity of the shared model or compromise the privacy of individual participants' data. To ensure the viability and reliability of FL, it is imperative to comprehensively identify, categorize, and address these security and privacy issues. 

In response to this need, this paper comprehensively explores security and privacy challenges in the context of Federated Learning. The main goal is to provide a comprehensive taxonomy of potential risks that may arise from both the aggregator and participating entities. By categorizing attacks into distinct classes, including poisoning attacks\cite{pmlr-v108-bagdasaryan20a}, backdoor attacks\cite{8835365}, membership inference attacks\cite{suri2022subject}, generative adversarial network (GAN) based attacks\cite{zhang2019poisoning}, and differential privacy attacks\cite{el2022differential}, we aim to shed light on the diverse array of threats faced by FL systems. The scope of this study encompasses various machine learning models, spanning from conventional algorithms to cutting-edge large language models. As FL finds applicability in various domains, such as healthcare, finance, and the Internet of Things (IoT), understanding and addressing the unique security and privacy challenges becomes even more crucial. As part of this investigation, we delve into the methodologies employed by malicious entities to compromise FL systems and intrude upon the privacy of participants' data. Furthermore, we discuss primary mitigation techniques that have shown promise in countering these security risks and upholding data confidentiality. Notably, we explore the integration of blockchain and Trusted Execution Environments as potential solutions to reinforce the security of FL systems.

This paper endeavors to present a comprehensive overview of the security and privacy landscape of Federated Learning. By identifying existing threats and highlighting future research directions, we aim to contribute to the ongoing efforts to fortify FL systems against emerging security risks and maintain the utmost protection of sensitive data in distributed learning environments. As FL continues to evolve, this study seeks to foster a more secure and privacy-conscious foundation for this promising approach to machine learning.

This paper is organized as follows: Section 2 delves into the fundamentals of Federated Learning (FL) and its underlying mechanism, ensuring data privacy and confidentiality. Moving on to Section 3, we explore the various Security and Privacy Challenges associated with FL, analyzing numerous potential attacks on this novel approach. Section 4 presents potential solutions to counteract these attacks and discusses defensive measures to bolster FL's security and privacy. Section 6 explores related works in this domain, shedding light on previous research and advancements. Finally, in Section 7, we draw the paper to a conclusion, outlining the future directions for enhancing security and privacy in Federated Learning.

% === II. Background ========================
% =================================================================================
\section{Background}

% \begin{equation}\label{ideal_rectifier_resistance}
% R(v) =
% \begin{cases}
%     \infty, & v > 0\\
%     0, & v \leq 0
% \end{cases}
% \end{equation}

% \begin{equation}\label{nonideal_rectifier_resistance}
% R(v) =
% \begin{cases}
%     \infty, & v > -V_{tr}\\
%     R_{on}, & v \leq -V_{tr}
% \end{cases}
% \end{equation}

\subsection{Federated Learning Concepts}
Federated Learning is a decentralized machine learning paradigm that facilitates training models across multiple devices while keeping the data on those devices, ensuring user privacy. Instead of sending raw data to a central server for training, FL allows devices, such as smartphones, edge servers, or Internet of Things (IoT) devices, to collaboratively learn from local data while keeping the data localized and secure.

\begin{figure}[!ht]
\centering
\includegraphics[width=3.5in]{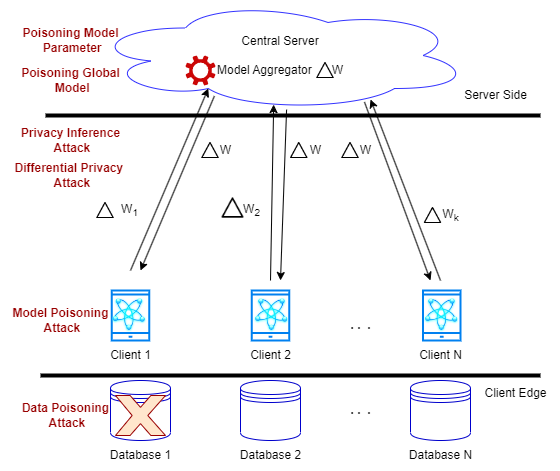}
\caption{Various Attack Models Within the FL Framework}
\label{fig_sim}
\end{figure}

\subsection {Data Privacy and Confidentiality}
One of the primary motivations behind Federated Learning is to preserve data privacy and confidentiality. In traditional centralized machine learning, sensitive data is often collected and stored on a central server, raising concerns about unauthorized access and potential data breaches. In FL, data remains on the devices where it originates, and only model updates or aggregated information is transmitted to the central server. This approach significantly reduces the risk of exposing raw, sensitive data to potential adversaries.

\subsection {Security and Privacy Challenges of FL}
Federated Learning presents a novel approach to training models while preserving user privacy, but it also introduces several security and privacy challenges that demand attention. In this section, we discuss some of the critical challenges FL faces, shown in Figure 1:

\textbf{Poisoning Attacks:} Poisoning attacks, also known as data poisoning attacks, involve adversaries injecting malicious data into the local training datasets of participating devices. These adversarial samples can skew the model's learning process, leading to biased or compromised global models when the updates are aggregated. Robust defenses are essential to detect and mitigate the impact of poisoning attacks on FL.

\textbf{Backdoor Attacks:} Backdoor attacks aim to create a "backdoor" in the model, allowing an attacker to trigger specific behavior or misclassification when presented with specific input patterns. These backdoors are often injected during training and can pose a significant threat, particularly in scenarios where models are shared across multiple devices and users.

\textbf{Membership Inference Attacks:} Membership inference attacks focus on inferring whether specific data samples were part of the training dataset used to create the global model. Successful membership inference attacks can compromise user privacy by revealing sensitive information about the data contributors. Developing robust mechanisms to prevent such inference is crucial for maintaining data privacy.

\textbf{Generative Adversarial Network (GAN) Based Attacks:} Generative Adversarial Networks (GANs) can be leveraged by adversaries to generate synthetic data that closely mimics real data distribution. These synthetic samples may then attack the FL system, potentially leading to data leakage or model manipulation. Detecting and countering GAN-based attacks is a critical challenge.

\textbf{Differential Privacy Attacks:} Differential privacy is a key technique used to protect individual data privacy in FL. However, FL systems are not immune to differential privacy attacks, where attackers attempt to reverse-engineer the presence of specific data points or learn sensitive information from the differentially private model updates. Strengthening the privacy guarantees against such attacks is essential.

Addressing these security and privacy challenges is vital to ensure the success and widespread adoption of Federated Learning. Robust defense mechanisms, privacy-preserving techniques, and continuous research efforts are needed to enhance the security posture of FL systems and protect user data and privacy. As we move forward, exploring innovative solutions and adopting a proactive approach to security will be instrumental in making FL a reliable and privacy-conscious framework for collaborative machine learning.

% === III. Security and Privacy Challenges =======================================
% =================================================================================
\section{Security and Privacy Challenges}

\subsection{Poisoning Attacks}
A significant and concerning attack prevalent in Federated Learning (FL) context is known as poisoning\cite{pmlr-v108-bagdasaryan20a}. Due to the decentralized nature of FL, where each client possesses its training data, the risk of incorporating tampered data weights into the global ML model becomes substantial. This poisoning attack can occur during the training phase and potentially impact both the local models and, consequently, the overall performance and accuracy of the global ML model. In FL, model updates are aggregated from a large group of clients, making the probability of poisoning attacks from one or more clients' training data quite high. Consequently, the threat posed by poisoning attacks is severe. These attacks specifically target various stages and components within the FL process. Below, we provide a concise overview of the different classifications of poisoning attacks: 

\textbf{Data Poisoning Attack:}
The inception of data poisoning attacks against machine learning algorithms dates back to the seminal work of \cite{biggio2012poisoning}, where the researchers introduced exploiting the vulnerabilities of support vector machines by incorporating malicious data points during the training phase to maximize classification errors.

Since then, various approaches have been proposed to counter data poisoning attacks in machine learning algorithms under various settings, including centralized and distributed environments. In Federated Learning (FL), where clients actively participate in the training process by contributing data and sending model parameters to the server, the risk of malicious clients poisoning the global model becomes evident. Data poisoning in FL refers to generating tainted samples to train the global model, aiming to produce falsified model parameters and transmit them to the server.

Another related aspect is data injection, which can be considered a subcategory of data poisoning. In this scenario, a malicious client may inject tainted data into the local model processing, potentially gaining control over multiple clients' local models and manipulating the global model with their maliciously crafted data. These works collectively highlight the importance of developing robust defense mechanisms to safeguard against data poisoning attacks in the ever-evolving landscape of machine learning algorithms and distributed learning settings.

Figure 2 demonstrates the impact of malicious clients on a CNN system involving 15 clients. These malicious clients upload fake parameter values during each communication round, which can be either opposite to the true value or random numbers within [-1, 1]. The results reveal that malicious clients adversely affect the system's performance. Moreover, as malicious clients increase, the system's reliability significantly diminishes, eventually leading to failure.

\textbf{Model Poisoning Attack:}
In model poisoning attacks\cite{fang2020local}, the malicious party can modify the model updates directly before sending them to the central server for aggregation. This enables them to inject malicious parameters into the global model, poisoning its integrity and functionality. The larger the scale of the FL system with numerous clients, the higher the potential effectiveness of model poisoning attacks.

\begin{figure}[!ht]
\centering
\includegraphics[width=3.5in]{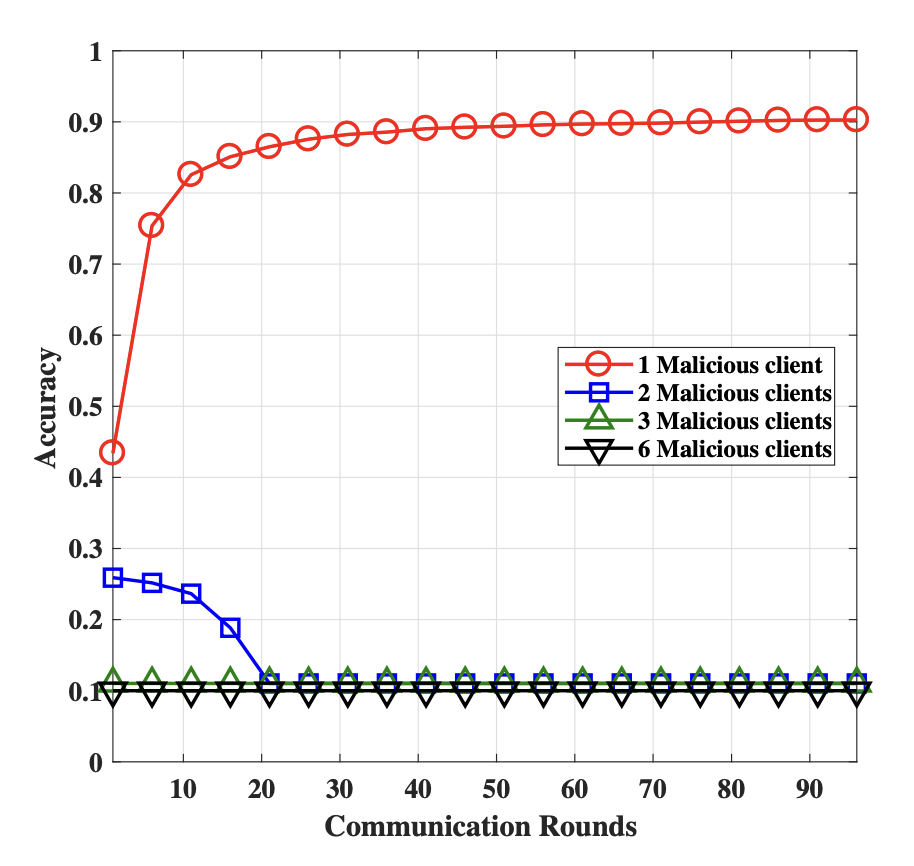}
\caption{Performance Comparison with Varying Number of Malicious Clients.}
\label{fig_sim}
\end{figure}

\subsection{Backdoor Attacks}
In machine learning security and privacy, transparent attacks like poisoning and inference attacks are known entities. However, lurking in the shadows is a more insidious threat known as backdoor attacks. Unlike their transparent counterparts, backdoor attacks cleverly inject a malicious task into an existing model while preserving its accuracy for the genuine task. This cloak-and-dagger approach makes them difficult to detect promptly, as they may not immediately impact the performance of the original ML task.

In\cite{sun2019can}, the authors experiment with and demonstrate the implementation of backdoor attacks. To mitigate these risks,\cite{liu2018fine} propose model pruning and fine-tuning as potential solutions. However, the severity of backdoor attacks remains high, as their occurrence often goes unnoticed for significant periods, allowing the attacker to maintain covert control. Backdoor attacks can significantly confuse ML models and confidently predict false positives. Trojan threats\cite{Xie2020DBA:} represent a similar category of backdoor attacks, where the attacker aims to maintain the ML model's primary task while performing malicious actions in stealth mode.

In federated learning, backdoor attacks have emerged as potential security threats. The main objective of a backdoor attack in FL is to manipulate local models to compromise the global model. In such attacks, the attacker introduces triggers into one or more local models, causing the global model to exhibit specific behaviors under the presence of these triggers in the inputs. For instance, in autonomous driving\cite{xie2021crfl}, an attacker may deploy a backdoored street sign detector that excels at identifying street signs under normal conditions but erroneously identifies stop signs with specific stickers as speed limit signs.

\subsection{Membership Inference Attacks}
Membership Inference attacks\cite{2018arXiv180709173T} aims to extract information by determining if specific data points are present in a model's training set. The attacker exploits the global model to gain insights into the training data of other users, potentially compromising their privacy and security.

\textbf{Training Data Inference Attacks:} Membership inference attacks refer to techniques that attempt to deduce details about the training data of a machine learning model. By exploiting the global model, attackers seek to ascertain whether specific data points were used during the model's training. These attacks rely on guesswork and training a predictive model to infer the original training data.

\textbf{Inference Attacks on Training Data:} These attacks uncover information about the training data used to build a machine-learning model. By manipulating the global model, attackers aim to determine the presence or absence of certain data points in the training set. Employing various techniques, they construct predictive models to make educated guesses about the original training data. 

\textbf{Training Data Reconnaissance Attacks:} Membership inference attacks are a form of survey aimed at gaining insights into the training data of a machine learning model. Exploiting the global model, attackers attempt to discern whether specific data instances were part of the model's training set. By training their predictive models, attackers employ educated guesswork to infer details about the original training data. Researchers have demonstrated the potential risks of memorizing neural networks' training data, exposing them to passive and active inference attacks.

\textbf{Data Set Inference via Model Exploitation:} These attacks involve exploiting a machine learning model to infer details about the training data. Attackers use the global model to check for the presence of specific data in the training set. By training their predictive models, they attempt to deduce information about the original training data through educated guesses. 

\textbf{Training Data Guessing Attacks:} Membership inference attacks are akin to "guessing" the training data used to train a machine learning model. By leveraging the global model, attackers attempt to deduce whether particular data points were present in the training set. Employing various predictive modeling techniques, they try to infer details about the original training data through educated guesses.

\subsection{Generative Adversarial Network (GAN) Based Attacks}
In the context of security and privacy challenges, the emergence of Generative Adversarial Networks (GANs) poses a new and potent threat. GANs, a powerful development in Deep Learning\cite{goodfellow2014generative}, continues to be actively researched and refined\cite{hitaj2017deep}. Their primary objective is to generate synthetic samples that closely resemble the distribution of the original training data, even without direct access to the original samples.

The GAN framework sets up a competitive game between two deep learning networks: generative and discriminative networks, akin to game theory. The generative network produces realistic samples, while the discriminative network aims to differentiate between real data samples and those generated by the GAN.
Initially applied to image datasets, GAN attacks have since shown a potential to be extended to diverse types of data, including sensitive records like demographic data. This raises significant concerns regarding privacy and security implications. Malicious agents can exploit GANs to create synthetic data that resembles genuine data, possibly leading to privacy breaches, data falsification, and adversarial manipulation of machine learning models.

\textbf{Attacks on Client Edge:}
The GAN-generated samples aim to closely imitate the distribution of the original training data. Applying record-level differential privacy noise, a technique previously suggested for privacy protection proves ineffective against GAN-based attacks\cite{shokri2015privacy}. 

The attack primarily relies on an active insider who operates under a white-box access model, gaining access to and using internal model parameters. The attacker participates in the federated deep learning protocol as an honest client but endeavors to extract information about a class of data they do not own (owned by the victim client). Through this active attack, the adversary influences the learning process to force the victim into releasing further details about the targeted class.
However, client-side GAN-based attacks have three main limitations: first, they require altering the distributed model's architecture to introduce adversarial influence in the learning process; second, the adversarial influence introduced by the malicious client may become insignificant after several iterations of the process; and third, the attack can only imitate input data for training rather than replicating exact samples from the victim side\cite{8835269}.

\subsection{Differential Privacy Attacks}
Security and privacy challenges in federated learning have prompted the widespread adoption of Differential Privacy (DP), a popular technique in industry and academia. DP's core concept revolves around preserving individual privacy by introducing noise to sensitive attributes, ensuring each user's data remains protected. Despite the addition of noise, the loss of statistical data quality is relatively low compared to the enhanced privacy protection. In federated learning, DP is a crucial defense against inverse data retrieval. By applying DP to participants' uploaded parameters, frameworks like DPGAN\cite{xie2018differentially} and DPFedAvgGAN\cite{augenstein2019generative} render GAN-based attacks inefficient in inferring other users' training data within the deep learning network. DP is versatile and finds application in various scenarios, as demonstrated in multi-agent systems, reinforcement learning, transfer learning, and distributed machine learning\cite{zhu2020more}.

Some works combine secure multiparty computation and differential privacy to achieve a secured federated learning model with high accuracy\cite{truex2019hybrid}. Additionally, other approaches improve privacy guarantees by combining DP with shuffling techniques and user data masking using an invisibility cloak algorithm\cite{ghazi2019scalable}. However, these solutions introduce uncertainty into the uploaded parameters, potentially compromising training performance. The challenge lies in balancing robust privacy protection and maintaining optimal training performance in federated learning systems. Developing techniques that effectively protect user privacy while preserving model accuracy and server evaluability remains an ongoing area of research to ensure the trustworthiness and reliability of federated learning frameworks.

\section{Possible Solutions}

\subsection{Defense Against Poisoning Attacks}
Poisoning attacks involve injecting malicious data into a machine learning model's training set to manipulate the model's behavior during training or deployment. Defense strategies against poisoning attacks typically involve data sanitization techniques, outlier detection, or verification mechanisms. Some popular defense methods\cite{osti_10294585} are Byzantine robust aggregation, clustering-based detection, and behavior-based detection methods to enhance the security and robustness of FL systems.

\subsection{Defense Against Backdoor Attacks}
Backdoor attacks involve adding a hidden pattern or trigger to a machine-learning model that causes it to produce incorrect results when triggered by specific inputs. Defense against backdoor attacks often involves model inspection, identifying and removing suspicious patterns, or techniques like fine-tuning to retrain the model and its parameter without the backdoor trigger.

\subsection{Defense Against Membership Inference Attacks}
Membership inference attacks attempt to determine whether a specific data point was used in a machine learning model's training dataset. Defending against such attacks may involve differential privacy techniques, adding noise to the training data, or employing privacy-preserving algorithms. These defense mechanisms are designed to protect the privacy and confidentiality of individual data points within the federated learning setting, thereby enhancing the security and privacy of the overall FL system.

\subsection{Defense Against Generative Adversarial Network Based Attacks}
GANs can generate realistic synthetic data, which could be misused to attack machine learning models. Defense strategies against GAN-based attacks might involve:
\begin{itemize}
    \item Adversarial training
    \item Utilizing GANs for data augmentation
    \item Employing detection mechanisms to identify fake data
\end{itemize}

\subsection{Defense Against Differential Privacy Attacks}
Differential privacy attacks attempt to infer sensitive information about individuals from a trained model. Defense against such attacks often involves incorporating differential privacy mechanisms during the training process to ensure the privacy of individuals' data. Some key strategies that can be followed to defend from such attacks include robust identity verification, formal methods, federated averaging with differential privacy, client selections, Byzantine fault tolerance, homomorphic encryption, transfer learning, model distillation, and secure model aggregations.

% ===================================================================================================================================
% ===================================================================================================================================

%
%\begin{figure}[!t]
%\centering
%\includegraphics[width=2.5in]{myfigure}
% where an .eps filename suffix will be assumed under latex, 
% and a .pdf suffix will be assumed for pdflatex; or what has been declared
% via \DeclareGraphicsExtensions.
%\caption{Simulation Results}
%\label{fig_sim}
%\end{figure}

%
% \begin{figure*}[!t]
% \centerline{\subfloat[Case I]\includegraphics[width=2.5in]{subfigcase1}%
% \label{fig_first_case}}
% \hfil
% \subfloat[Case II]{\includegraphics[width=2.5in]{subfigcase2}%
% \label{fig_second_case}}}
% \caption{Simulation results}
% \label{fig_sim}
% \end{figure*}
%

\section{Related Work}
This section reviews existing research and studies related to Federated Learning (FL), specifically focusing on security and privacy aspects. The landscape of FL research has grown rapidly, and numerous studies have contributed valuable insights into addressing the challenges associated with this decentralized learning paradigm.

\textbf{Privacy-Preserving FL Techniques:} Several research efforts have explored novel privacy-preserving techniques in FL. Differential privacy has been a prominent approach to protect individual data privacy during the aggregation of model updates. Various studies have proposed improved differential privacy mechanisms tailored for FL settings, ensuring a balance between privacy guarantees and model accuracy.

\textbf{Adversarial Attacks and Defenses:} Research on adversarial attacks and defenses in FL has gained significant attention. Studies have investigated poisoning attacks, backdoor attacks, membership inference attacks, and other adversarial strategies to compromise FL systems. Researchers have proposed robust defenses to counteract these threats, such as anomaly detection, secure aggregation protocols, and adversarial training mechanisms.

\textbf{Secure Communication Protocols:} Communication security is critical to FL, as model updates are transmitted between devices and the central server. Several studies have focused on developing secure communication protocols, incorporating encryption, authentication, and integrity verification techniques to safeguard against eavesdropping and tampering.

\section{Conclusion}
In summary, Federated Learning (FL) has emerged as a promising paradigm to address the challenges of centralized data storage and privacy concerns in machine learning. By decentralizing data and enabling collaborative learning across multiple devices, FL offers a novel approach that preserves user privacy while harnessing the power of AI across diverse domains. Throughout this paper, we have explored the fundamental concepts of Federated Learning, delving into its underlying workings for data privacy and confidentiality. We also identified and discussed FL's various security and privacy challenges, including poisoning attacks, backdoor attacks, membership inference attacks, GAN-based attacks, and differential privacy attacks.

Researchers and practitioners must develop robust defense mechanisms, secure communication protocols, and privacy-preserving techniques to address these challenges. Techniques such as differential privacy, anomaly detection, and adversarial training can be vital in safeguarding FL systems against adversarial threats and data breaches. As FL gains traction and finds applications in real-world scenarios, regulatory and ethical considerations become paramount. Adherence to data protection regulations, informed consent, and ethical data handling are essential to maintain public trust and confidence in FL technologies.

% ============================================
\ifCLASSOPTIONcaptionsoff
  \newpage
\fi

% ====== REFERENCE SECTION

%\begin{thebibliography}{1}

% IEEEabrv,

\bibliographystyle{IEEEtran}
\bibliography{IEEEabrv,Bibliography}

\vfill

% that's all folks
\end{document}